\documentclass[aps,prd,onecolumn,nofootinbib,notitlepage]{revtex4-1}
\usepackage{epsfig}
\usepackage{amsmath}
\usepackage{amsfonts}
\usepackage{amssymb}
\usepackage{graphicx}
\usepackage{colordvi}
\usepackage{float}
\usepackage{colortbl}
\usepackage{bm}
\usepackage{latexsym}
\usepackage{psfrag}
\usepackage{amsthm}
\usepackage{color}
\usepackage{braket}
\usepackage{indent first}
\usepackage[normalem]{ulem}
\usepackage{fancyhdr}
\usepackage{hyperref}
\usepackage{environ}
\usepackage{mathtools}
\usepackage{float}
\usepackage{comment}
\usepackage{subcaption}
\newcommand{\be}{\begin{equation}}
\newcommand{\ee}{\end{equation}}
\newcommand{\bea}{\begin{eqnarray}}
\newcommand{\eea}{\end{eqnarray}}
\begin{document}
%
%
%
%
%
%
\title{ Effects of   underlying  topology  on quantum state  discrimination    }
\author{Aatif Kaisar Khan}
\email{sahbran8@gmail.com}
\affiliation{Perimeter Institute for Theoretical Physics, Waterloo, Ontario N2L 2Y5, Canada}

\author{Yasir Hassan Dar}
\email{yasirhassan285@gmail.com}
\affiliation{Perimeter Institute for Theoretical Physics, Waterloo, Ontario N2L 2Y5, Canada}

 \author{Elias C. Vagenas}
\email{elias.vagenas@ku.edu.kw }
\affiliation{ Department of Physics, College of Science, Kuwait University, Sabah Al Salem University City, P.O. Box 2544, Safat 1320, Kuwait}
\author{Salman Sajad Wani}
\email{sawa54992@hbku.edu.qa}
	\affiliation{Qatar Center for Quantum  Computing, College of Science and Engineering, \\Hamad Bin Khalifa University, Qatar}
 \author{Saif Al-Kuwari}
\email{smalkuwari@hbku.edu.qa}
\affiliation{Qatar Center for Quantum  Computing, College of Science and Engineering, \\Hamad Bin Khalifa University, Qatar}

\author{Mir Faizal}
\affiliation{Canadian Quantum Research Center 204-3002, 32 Ave Vernon, BC V1T 2L7 Canada}
\affiliation{Irving K. Barber School of Arts and Sciences, University of British Columbia Okanagan Campus, Kelowna, BC V1V1V7, Canada}
\email{mir.faizal@ubc.ca}

\begin{abstract}
\par\noindent
In this work, we show that quantum state discrimination can be modified due to a change in the underlying topology of a system.   In particular, we explicitly demonstrate that the quantum state discrimination of systems with underlying discrete topology differs from that of systems with underlying continuous topology. Such changes in the topology of a spacetime can occur in certain quantum gravity approaches. In fact,  all approaches to quantum gravity can be classified into two types: those with underlying continuous topology (such as string theory) and those with an underlying discrete topology (such as loop quantum gravity). We demonstrate that the topology of these two types of quantum gravity approaches has different effects on the quantum state discrimination of low-energy quantum systems. We also show that any modification of quantum mechanics, which does not change the underlying topology, does not modify quantum state discrimination.  
\end{abstract}
\maketitle
%
%
%
%
\par\noindent
In quantum mechanics, any physical quantity is calculated using the quantum state of the system. However, to perform such calculations (and henceforth, measurements based on the results of such calculations), we need to identify the quantum state of the system. Performing a small number of measurements on the system to obtain its quantum state is called quantum state discrimination \cite{5a, 5b}. Although quantum state discrimination has been used in important physical settings, such as noise-robust quantum networks \cite{a0},   noisy quantum neural networks \cite{a1},  quantitative wave-particle duality \cite{a2}, decomposable entanglement witness \cite{a3}, entanglement swapping \cite{a4}, and no-signaling bound \cite{a5}, the effect of the underlying topology of the system on the quantum state discrimination has never been studied. Therefore, in this paper,  we analyze the effect of the underlying topology on the quantum state discrimination and deduce that the underlying topology of the system can non-trivially modify the quantum state discrimination.  This observation can have important implications for quantum gravity phenomenology.


The foundations of modern physics are generally based on two important theories: quantum mechanics and relativity. Although quantum mechanics and special relativity can be reconciled in the framework of quantum field theories \cite{1}, this is not true for general relativity and quantum mechanics \cite{2}. Therefore, in the absence of a complete and self-consistent theory of quantum gravity, various approaches have been developed, such as string theory \cite{3}, loop quantum gravity \cite{4}, Euclidean quantum gravity \cite{5}, Wheeler-DeWitt approach \cite{6}, spin foam \cite{7}, and causal sets \cite{8}. 
While the underlying topology of a spacetime is taken as continuous, there are indications that this could only be an approximation. Thus, all these approaches can be broadly classified into two main classes. In the first class, the spacetime emerges from a discrete structure, such as loop quantum gravity  \cite{4} and in the second class, the spacetime is continuous, such as string theory \cite{3}. We emphasize that the spacetime structure could still have an intrinsic cutoff, in the form of a minimal length, but it is not discrete spacetime. There are ways to incorporate a cutoff in the form of a minimal length or even a minimal time without making the spacetime topology discrete. This is done in the context of string theory using T-duality \cite{9}. Due to T-duality, there is a minimal scale for gaining new information beyond which no new information is obtained.  Hence, string theory cannot be probed below the minimal scale. This scale is a sophisticated way to incorporate the naturally expected result that string theory cannot be probed below. This is also the case in perturbative string theory, where a fundamental string is the only quantum object available, and it is impossible to probe spacetime below the length of this fundamental string \cite{10}. However, due to T-duality, these results also hold when non-perturbative point-like objects, such as D0-branes, are considered \cite{11}.  At this point, it is worth of noting that  it has been possible to study, the T-duality in a covariant formulation of string theory \cite{ref1, ref2, ref3}. In this formalism, the properties of string theory can be related to the foundations of quantum theory in terms of modular variables. In addition, it has been demonstrated that it can help understand aspects of quantum non-locality and it is also consistent with causality \cite{ref4}. 

In this context, the principle of relative locality has been proposed to construct a spacetime with minimal length but without a discrete topology \cite{1b}. The main difference between a cutoff in the form of discrete spacetime and a minimal length is that, in discrete spacetime, the spacetime emerges from a discrete lattice. The lattice points are fixed, and the emergent structure appears as a continuum at a scale large enough to neglect this underlying discrete topology. On the other hand, it is not possible to define any short-distance structure from which spacetime emerges in other more sophisticated approaches to quantum gravity that have an intrinsic minimal length. Thus, spacetime can be measured or even meaningfully defined, only up to a certain length scale. It is evident that this is a rather small difference, but it does produce different types of corrections to low-energy quantum systems. These corrections can then be used to discriminate between these two classes of quantum gravity approaches. 
Although, at present, we do not know which approach to quantum gravity is the correct one, here we propose that we can observe completely different effects on the quantum state discrimination of these two classes. Both these classes have low-energy implications; however, the class of quantum gravity approaches with continuous spacetime topology produces a quartic correction to the low-energy effective field theory describing simple quantum systems \cite{1c, 1c1, 1c2, 1c4}, whereas the class of quantum gravity approaches with discrete topology produces a cubic correction to the low-energy effective field theory describing simple quantum systems \cite{1d, 1d1, 1d2, 1d4,Khan:2022yfm}.
In fact, these cubic corrections occur in any quantum system with underlying discrete topology as the leading order corrections of the effective theory are considered. 
They have even been observed in effective field theory describing condensed matter systems like graphene, due to the underlying discrete structure of such condensed matter systems \cite{grp1, grp2}. Similarly, the quartic corrections occur in any quantum system with underlying continuous topology as the leading order correction of the effective field theory, such as non-linear optics \cite{eff1, eff2}.
This observation can be used to critically modify the quantum state discrimination.  Thus, quantum state discrimination will be different for theories whose leading order correction term is a cubic correction term compared to those whose leading order correction term is quartic.
Several works have been proposed to test the effects of both quartic and cubic corrections on quantum systems, and various tabletop experiments have been proposed to detect such corrections \cite{2a, 2b, 2c, 2d}. Despite the fact that these corrections have not been detected to date, such a phenomenological approach to quantum gravity has been used to propose bounds on the scale at which quantum gravitational effects become important. Here, we do expect quantum gravitational effects to become important at the Planck scale. However, no rigorous arguments prevent such effects from becoming important at scales much larger than the Planck scale \cite{4a, 4b}. Various experiments have been proposed to test these effects, but this is the first time that quantum state discrimination has been directly related to various quantum gravitational approaches. In particular, we present a novel method that can be used to at least show which class of quantum gravity approaches actually ``represents" quantum gravity. Since string theory and loop quantum gravity are placed in different classes, we can at least deduce which of these two theories can ``represent" quantum gravity. However, we point out that quantum state discrimination cannot be used to separate two quantum gravity approaches that both have an underlying discrete structure or two quantum gravity approaches with spacetime that is not discrete at any scale. Hence, for example,  the quantum state discrimination will not be modified between the causal set approach and loop quantum gravity, or between string theory and Euclidean quantum gravity. However, the subtle observation that quantum state discrimination changes between a wide class of quantum gravity approaches is a novel result and can help determine the future direction of research in quantum gravity.\\
%
%
%
%
%
%
%
%
%
%
%
%
%
%
%
\par\noindent
We start by analyzing the corrections to the standard quantum mechanics from quantum gravity.
These quantum gravity corrections depend on the underlying topology of the theory and, thus, they will be different for different quantum gravity approaches. 
It is known that quantum gravity approaches, such as string theory, modify the Heisenberg algebra, and this modified Heisenberg algebra produces a quartic correction term in the  Schr\"odinger equation \cite{1c, 1c1, 1c2,  1c4}. However, quantum gravity approaches, such as loop quantum gravity,   modify the Heisenberg algebra in such a way that the Schr\"odinger equation is modified by a cubic correction term \cite{1d, 1d1, 1d2,  1d4}. 
Here, we will include both quartic \cite{1c, 1c1, 1c2,  1c4} and cubic \cite{1d, 1d1, 1d2,  1d4} correction terms  to the Schr\"odinger equation, and, thus, we will write it in the form:
\begin{equation}
\biggr(  -\frac{\hbar^2}{2m}\frac{d^2}{dx^2}- \frac{i\alpha\hbar^3}{m}\frac{d^3}{dx^3}+ \frac{\beta \hbar^4}{m} \frac{d^4}{dx^4}  +V(x) \biggl)\psi(x)=E\psi(x)
\label{t1}
\end{equation}
where $\alpha$ and $\beta$ parameterize the scales at which such corrections occur. These parameters are fixed by current experimental data  \cite{2a, 2b, 2c, 2d}. As stated, the cubic corrections occur due to the underlying discrete topology and, thus, have been observed as higher order corrections to certain condensed matter systems \cite{grp1, grp2}. Similarly, quartic corrections occur for systems with underlying continuous topology, such as non-linear optics \cite{eff1, eff2}. Hence, even though these corrections were initially motivated by quantum gravity approaches \cite{1c, 1c1, 1c2,  1c4, 1d, 1d1, 1d2,  1d4}, we keep our discussion general, as they are not limited to quantum gravity.  In fact,  we consider both of them for completeness, which allows us to analyze their relative effects on the quantum system. 
 At this point, we should point out that in the framework of the statistical mechanics of D0-branes, a similar to Eq. (\ref{t1}) expansion  with higher derivative corrections in momenta was obtained \cite{ref8, ref8a, ref8b, ref8c, ref8d}. As expected (due to the spacetime being continuous in this model), the leading order correction  is a quartic correction term. In this model, explicit expansion has been obtained for higher powers of momentum, and it has been observed that this model contains only even powers of momentum. 
%
%
%
%
\par\noindent

To use quantum state discrimination \cite{5a, 5b}, we consider a Schr\"odinger equation with a double-delta well potential.  This double-delta potential can be used to realize several interesting physical systems, such as neutron-bound states \cite{n1}, PT-symmetric Bose-Einstein condensate \cite{n2}, and the Ramsauer-Townsend  effect \cite{n4}.  In these physical systems,  the effects of the modified quantum mechanical behavior on the quantum state discrimination can be tested. The specific potential $V(x)$ that represents a  double-delta well with the wells to be at $\pm L$, is given by (see Fig. 1)
\begin{equation}
V(x)=-W_o \, \delta (x+L) - W_o \,\delta (x-L)~.
\label{t2}
\end{equation}
\begin{figure}[h]
\includegraphics[scale=1]{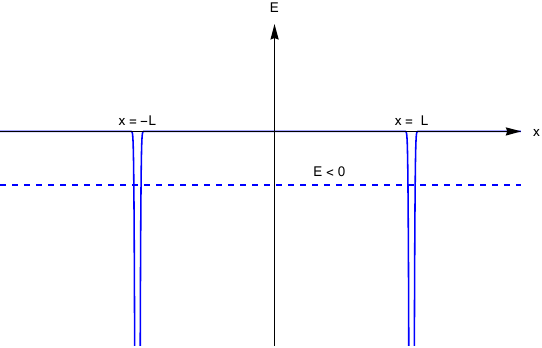}\label{a}\\
\caption{Double-Delta Potential Function}
\end{figure}
%
%
%
%
%
%
%
%
%
%
%
%
%
%
%
%
%
%
%
%
%
%
\par\noindent
The quantum state of a system can be used to estimate any observable needs to measure.  However, to use a quantum state to estimate any observable, the quantum state of the system itself has to be identified. In the absence of prior information, it is not possible to identify such a state. 
This is achieved by quantum state discrimination, which provides a way to distinguish between the different quantum states of a system. In quantum state discrimination, the discrimination error is minimized via a specific process that returns the minimum error in determining the quantum states \cite{h1, h2}. Other ways to approach quantum state discrimination exist, such as the maximum confidence, which gives the most likelihood of event detection in a measurement process. However, in this work, we will focus only on the minimum error discrimination for a two-state system \cite{h4} and consider the minimum error discrimination for a two-state double-delta well potential system. We start with the prior information that the quantum system is in one of the two states with associated probabilities of occurrence: $p_0$ and $p_1$, respectively. The probability of making an error in determining a state, $P_{Err}$, is  given by \cite{5a, 5b}
\begin{equation}
   P_{Err}= p_0- Tr\left( (p_0|\Psi_0^{\alpha}\rangle\langle\Psi_0^{\alpha}|)-p_1|\Psi_1^\alpha\rangle\langle\Psi_1^\alpha| \pi_0) \right)
\end{equation}
where, $p_0$ and $p_1$ denote the probability of occurrence of the states $|\Psi_0^\alpha>$ and $|\Psi_1^\alpha>$, respectively. In addition,   $p_0+p_1=1$ since there are only these two states.
Furthermore, we assume that for the probability operator $\pi_0$, the outcome is $0$ with the associated state being $\Psi_0^\alpha$ while  the probablility outcome for  $\pi_1$  is $1$  with the associated state being $\Psi_1^\alpha$. 
The minimum error for determining the state for this two-state system is given by the Helstorm bound \cite{h1, h2}  
%
%
%
%
%
%
\begin{equation}
P_{Min\, Err}=\frac{1}{2}(1-\sqrt{1-4p_0p_1|<\Psi^\alpha_0|\Psi^\alpha_1>|^2} )~.
\label{t48}
\end{equation}
\par\noindent
In the case of the double-delta potential, these two wavefunctions represent the two energy eigenfunctions, i.e, $|\psi^{\alpha}_{upper}>$ and $|\psi^{\alpha}_{lower}>$. 
In general, for the standard  Schr\"odinger equation, $|\Psi^{\alpha=0}_0>$ and $|\Psi^{\alpha=0}_1>$ can be represented as
$ 
|\Psi^{\alpha=0}_0>=\cos\theta_0|\psi^{\alpha=0}_{upper}>+\sin\theta_0|\psi^{\alpha=0}_{lower}> $ and $
|\Psi^{\alpha=0}_1>=\cos\theta_1|\psi^{\alpha=0}_{upper}>-\sin\theta_1|\psi^{\alpha=0}_{lower}>.
$, repspectively.
Thus, we can write 
$ 
<\Psi^{\alpha=0}_0|\Psi^{\alpha=0}_1>=\cos(\theta_0+\theta_1)
$.
Using these expressions, we obtain 
$  
P_{Min\, Err}^{\alpha=0}=(1-\sqrt{1-4p_0p_1\cos^2(\theta_0+\theta_1)})/2 
$.

Despite the fact that quantum state discrimination has been used to address various important physical problems, such as noise-robust quantum networks \cite{a0},   noisy quantum neural networks \cite{a1},  quantitative wave-particle duality \cite{a2}, decomposable entanglement witness \cite{a3},
entanglement swapping \cite{a4},  and no-signaling bound \cite{a5}, the effect of the modification of quantum mechanics on quantum state discrimination has never been studied. Quantum state discrimination can be utilized to study effects due to quantum mechnical modifications on such systems.  In this work, we use the quantum discrimination that has been formally developed as a field \cite{5a, 5b} to analyze the effect of such correction terms. For the  Schr\"odinger equation corrected by cubic and quartic corrections terms, the wavefunction will be modified (see Appendix A for cubic and Appendix B for quartic correction terms). We let $|\Psi^{\alpha}_0>$ and $|\Psi^{\alpha}_1>$ to be general modified states given by
$ 
|\Psi^{\alpha}_0>=\cos\theta_0|\psi^{\alpha}_{upper}>+\sin\theta_0|\psi^{\alpha}_{lower}> 
$ and $
|\Psi^{\alpha}_1>=\cos\theta_1|\psi^{\alpha}_{upper}>-\sin\theta_1|\psi^{\alpha}_{lower}>
$. 
Thus, again we have
$ 
<\Psi^{\alpha}_0|\Psi^{\alpha}_1>=\cos(\theta_0+\theta_1)
$. 
Using the corrected Schr\"odinger equation (with both cubic and quartic corrections obtained in  Appendix A and B), we obtain 
 $
P_{Min\, Err}^{\alpha}= (1-\sqrt{1-4p_0p_1\cos^2(\theta_0+\theta_1)})/2
$. 
This expression is similar to the expressions obtained for the standard Schr\"odinger equation. Thus, a modification of the Schr\"odinger equation does not have any direct effect on quantum state discrimination. 
\par\noindent
It should be stressed that the cubic corrections are only consistent with the discrete underlying topology (see Appendix A), and a change  in the underlying topology can produce a modification to the quantum state discrimination. 
On the contrary, the quartic corrections do not produce such topological changes and the topology of spacetime remains continuous. However, here we observe the discrimination between the quantum states at shorter lengths between the wells compared to the cubic corrections or the standard case (see Appendix B). Therefore, for analyzing the effects of topology on quantum state discrimination, we neglect the quartic correction terms.
We note that cubic correction terms follow directly from the discrete underlying topology of spacetime. In fact, cubic correction terms in the low-energy effective field theory of any system imply that the underlying topology is discrete \cite{1d, 1d1, 1d2,  1d4, Khan:2022yfm}. This has been done by first observing that any physical length can in principle be represented by an infinite well (as the wavefunction of any particle used to probe such a length will vanish on its boundaries). Then, it was observed that any such length is a multiple of a fixed small quantized length if cubic corrections are considered. As this result was demonstrated to hold for any arbitrary length, we conclude that this could only be possible if any length (or even any physical quantity, which is expressed as a function of length) is discrete, and hence such cubic correction would require discrete underlying topology of spacetime. 
%
%
%
%
%
%
%
%
%
%
%
%
%
%
%
%
%
%
\par\noindent
Consequently, we set in Eq.  (\ref{t1})  the scale parameters to $\alpha\ne0$ and $\beta=0$. First, we generalize the results to a double-delta potential \cite{n1, n2, n4}, and, then, demonstrate that, like an infinite well, this potential also leads to discreteness. With $k\equiv \sqrt{{-2mE}/{\hbar^2}} $, 
and $V(x) =0$, we can write 
$ 
\psi(x)=Ae^{i\gamma x}+Be^{k_1 x}+Ce^{-k_2x}
$
where, $A$, $B$, and $C$ are constants,
$ 
\gamma\equiv {1}/{2\alpha\hbar},\quad k_1 \equiv k(1-i\alpha\hbar k)$, and $k_2 \equiv k(1+i\alpha\hbar k)
$. 
After imposing the bound state condition, namely $\psi(-\infty)=\psi(\infty)=0$, the wave-function becomes 
$\psi(x) = s_1e^{i\gamma x}+c_{11}e^{ k_1x}$ for $x< -L$, 
$\psi(x) = s_2e^{i\gamma x}+c_{21}e^{ k_1x} + c_{22}e^{-k_2x}$ for $x\in (-L,L)$ and 
$\psi(x) = s_3e^{i\gamma x}+c_{32}e^{-k_2x}$ for $x> L$
where, $s_1$, $s_2$, $s_3$, $c_{11}$, $c_{21}$, $c_{22}$, and $c_{32}$ are all constants of integration. We restrict  $c_{11}$ to real values (by eliminating any imaginary phase associated with it)  and, which implies further calculations. 
For the standard (unmodified) system, the $s_i$'s are all zero. This can also be seen as a necessary condition for the wavefunction to be physical. Let us now take  $s_1e^{i\gamma x}$  as an example to demonstrate this. By Taylor expansion, we obtain $s_i[ix/2\alpha \hbar)+-x^2/4\alpha^2 \hbar^2+O(1/\alpha^2)$]. Thus, as $\alpha\to 0$, it is obvious that $s_i$ must also vanish in order to have a well-defined wavefunction. This means that given $\alpha << 1 $, we must also have $s_i<<1$. Consequently, any term which has $s_i \alpha$ or any higher power of $s_i$ and $\alpha$ will vanish (this will be used to simplify expressions at various places, such as in Eq. (\ref{t27})).
Hence, we get
$ 
\lim_{\alpha\to 0}s_i= 0  \text{, for i=1,2,3}
$.
Therefore, for sufficiently small  $\alpha$, $s_i$ will also be small $
s_i\alpha^m= 0=s_i^n\quad \forall\,
 m\ge 1,\,
 n>1
$.  
To simplify the calculations, we define the following new quantities  $
      \bar{L}\equiv {mW_0L}/{\hbar^2}, \, \, 
      \bar{\gamma}\equiv \gamma L={L}/{2\alpha\hbar}, $ and $
      \bar{k}\equiv kL $. 
After fixing the arbitrary constants, the new length $\bar{L}$ now reads  (see Appendix A)
\begin{equation}
\bar{L}=\frac{\bar{k} \mp \frac{\bar{k}s_1}{c_{11}}\bigg[ \frac{2\bar{L}}{\bar{k}}e^{-(\bar{k}+i\bar{\gamma})} + \bigg(1-\frac{\bar{L}}{\bar{k}}\bigg)e^{\bar{k}} (e^{i\bar{\gamma}}\mp e^{-i\bar{\gamma}}) \bigg]}{(1\pm e^{-2\bar{k}})}~.
\label{t60}
\end{equation}
As the above equation originates in the quadratic equation solution, it will be represented by two different wavefunctions (which is also the case in the non-deformed Heisenberg algebra). The modifications produced in them are represented by the terms proportional to $s_1$. Here, $\bar{L}$ for the upper sign and the lower sign in the equation will be represented by $\bar{L}_u$ and $\bar{L}_l$, respectively.
Equating the imaginary parts of Eq. (\ref{t60}) and  that $s_1, c_{11}\in\mathbb{R}$, we find the condition, which holds  for both these wavefunctions: 
$
 L=2n\pi\alpha    \hbar
$ or $\bar{L}=2 n \pi \alpha m W_0/\hbar$, with $n\in \mathbb{N}$. 
Thus, any arbitrary length $L$ can only be a multiple of a fixed quantized value. Similar to the infinite well potential, we here, too, obtain the expected result, i.e., the cubic terms are produced by a discrete underlying topology.     
This   discrete nature of space is constructed by a quantum of length, given by 
$2\pi \alpha\hbar$
(similar results can be found in \cite{Das:2020ujn}).
We also observe that the quantum of length scales with $\alpha$, and, thus, it could only come from the cubic correction terms. Furthermore, there is no $\alpha^2$ contribution to the quantum of length, which means that there is no contribution from quartic correction terms; this is expected as such terms were motivated by theories with continuous underlying topology.

%
%
%
%
\par\noindent
We now consider the case where the scale parameters are set to $\alpha=0$ and $\beta\ne0$, and the potential is $V(x)=0$. Hence, we have $
\psi(x)=Ce^{\eta x}+D^{-\eta x}+Ae^{bx}+Be^{-bx}$, where $ \eta\equiv (\hbar\sqrt{2\beta})^{-1} , \, b\equiv k+\hbar^2k^3\beta$ and $A,\, B,\, C$, and $D$ are constants. After implementing the boundary conditions $\psi(-\infty)=\psi(\infty)=0$, we get, 
$\psi(x) = s_{11}e^{\eta x}+c_{11}e^{ bx}\,\text{for} \,  x< -L$, $\psi(x) =s_{21}e^{\eta x}+s_{22}e^{-\eta x}+c_{21}e^{ bx} + c_{22}e^{-bx}\,\text{for} \,  x\in (-L,L)$, and
$\psi(x) =s_{32}e^{-\eta x}+c_{32}e^{-bx}\, \text{for} \,   x>L$, where $s_{11},\,s_{21},\,s_{22},\,s_{32},\,c_{11},\,c_{21},\,c_{22},\, \text{and}\, c_{32} $ are constants of integration. Also, we have $s_{11}/c_{11} > 0$  (see Appendix B). In addition, similar to the $s_i$'s in the cubic correction case, here $s_{ij} \to 0$ as $\beta \to 0$. Therefore, when $\beta$ is small, $s_{ij}$'s will also be small. Thus, we have 
 $
s_{ij}\beta^m= 0=s_{ij}^{n}, \forall\,
 m\ge 1$ and  $n>1
$.  
Here, again, we obtain a condition that the length has to satisfy, which is given by (see Appendix B) 
\begin{equation}
     \bar{L} =   \frac{ \bar{k}-\lambda\frac{ \bar{k}^3}{ \bar{L}^2} - \frac{s_{11}}{c_{11}} \bar{k} e^{-\frac{ \bar{L}}{\sqrt{2\lambda}}+ \bar{k} + \lambda \frac{\bar{k}^3}{\bar{L}^2}} }{(1\pm e^{-2 (\bar{k} + \lambda \frac{\bar{k}^3}{\bar{L}^2})})}
     \label{g78}
\end{equation}
where $\lambda \equiv \beta m^2W_0^2/\hbar^2$. It is clear from this equation that everything here is real and, thus, consistent with the continuous underlying topology. Hence, we conclude that quartic correction terms are consistent with the continuous underlying topology of a spacetime. 
We now consider the minimum length between the wells, which is necessary for the existence of a particular wavefunction, represented by $\bar{L}_0$. In the standard (unmodified)  case and the case involving cubic correction terms, $\bar{L}_0$ is $0$ and $1/2$ for the wavefunction obtained by taking the upper sign and the lower sign, respectively, in Eq. (\ref{t60}) (see Appendix A). However, for the case where quartic correction terms are considered, we observe a difference, namely $\bar{L}_0$ is $0$ and $(1-(s_{11}/c_{11})\exp{({- \bar{L}_0/\sqrt{2\lambda}}}))/2$ for the wavefunction obtained by taking the upper sign and the lower sign, respectively, in Eq. (\ref{g78})  (see Appendix A). Hence, the effect of the quartic correction terms on the wavefunction   occurs before the effect of the quartic cubic correction terms on the wavefunction. 
\par\noindent
We now show that the discrete underlying topology directly modifies the quantum state discrimination. This can be observed by starting with 
 the probabilities 
$
p_0=p_1={1}/{2}
$, and analyzing the effect of the discrete underlying topology on them. 
Due to the  discrete underlying topology, we have $
P_{Min\, Err}=(1-|\sin(\theta_0+\theta_1)|)/2
$~.  This expression is valid only for discrete values of $\bar{L}$, which are  $\bar{L}=n ({2\pi \alpha mW_0}/{\hbar})$. This, along with the observation that  for the two states to exist, we need $2\bar{L}\ge 1$, 
leads to the condition (see Appendix A)
\begin{equation}
n \ge \bigg\lceil{\frac{\hbar}{4\pi\alpha m W_0}} \bigg\rceil \, \, \, \,\, \, \, \,   \text{ or } \, \, \, \, \, \, \, \,   L \ge 2\pi\alpha\hbar\bigg\lceil{\frac{\hbar}{4\pi\alpha m W_0}} \bigg\rceil~.
\end{equation}
This changes the behavior of the quantum state discrimination, which can be analyzed by plotting it. 
Consider, for our case, the minimum value of $n=3$ as depicted in Fig. 2. From these figures, it can easily be seen that the presence of $\alpha$ breaks the continuity of the plot, rendering it to be observable only at discrete values. We also note that the distance between the discrete points is directly related to $\alpha$. Hence, the quantum state discrimination is only affected by models of quantum gravity with discrete underlying spacetime topology. It does not get modified for a spacetime with continuous underlying topology, and, hence, a quantum gravity approach, such as string theory, will not change the behavior of the low-energy quantum state discrimination of quantum mechanics. On the contrary, a quantum gravity  approach, such as loop quantum gravity, can indeed affect it. Here, it is important to note that the original modification of the Heisenberg algebra did not directly effect the quantum state discrimination, but it was the discreteness of spacetime that modified it. Consequently, it would also be modified for any system with discrete underlying topology, leading to applications even beyond quantum gravity. 
%
%
%
%

\begin{figure}
    \begin{subfigure}[b]{0.35\textwidth}
        \includegraphics[width=\textwidth]
        {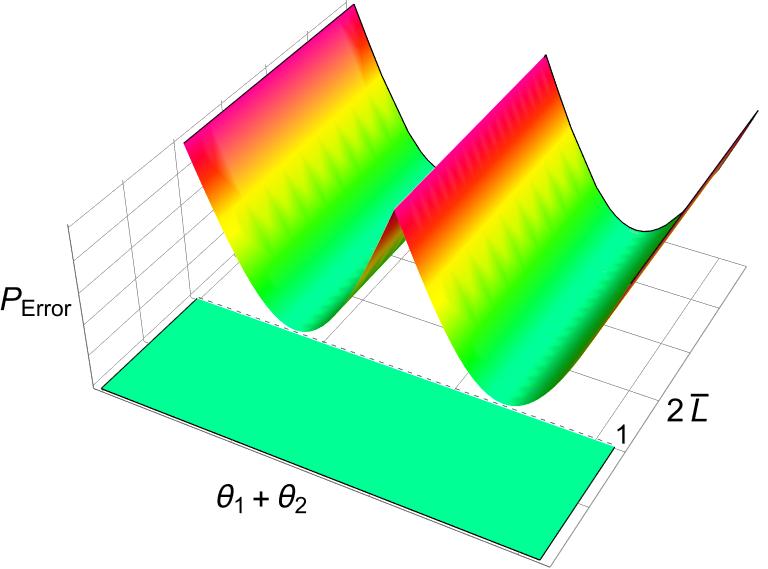}
        \caption{\ $\alpha = 0$ and $\beta=0$}
        \label{fig:image1}
    \end{subfigure}
    \hfill
 \begin{subfigure}[b]{1\textwidth}
        \includegraphics[width=\textwidth]
        {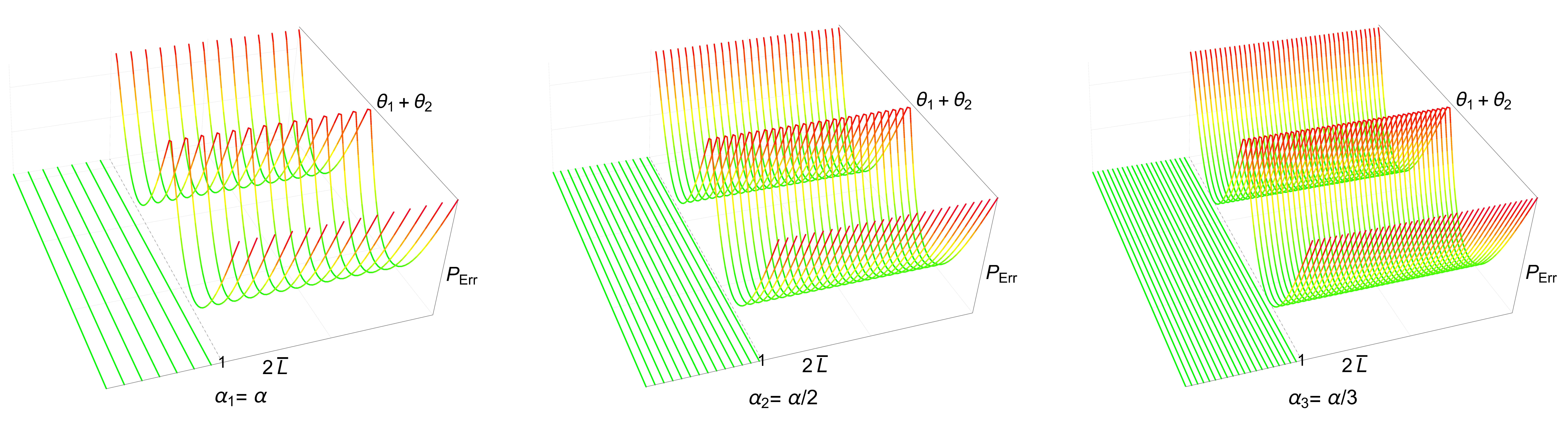}
        \caption{
         $\alpha\ne 0$ and $\beta = 0$
        }
        \label{fig:image2}
    \end{subfigure}
  
    \vfill
  
    \begin{subfigure}[b]{1\textwidth}
        \includegraphics[width=\textwidth]
        {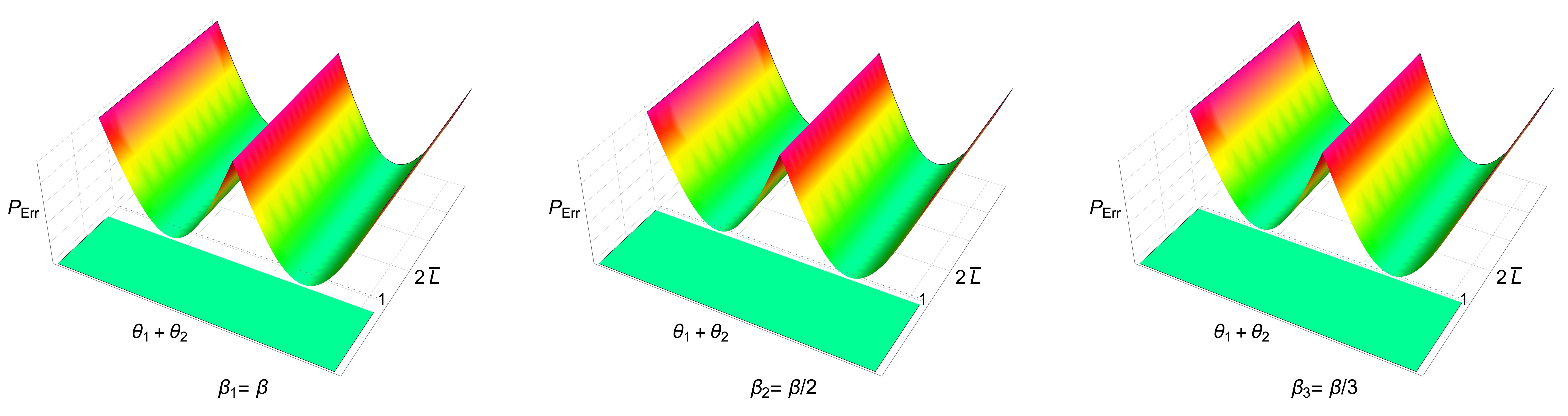}
        \caption{
        $\alpha=0$ and $\beta \ne 0$}
        \label{fig:image3}
    \end{subfigure}
    \hfill
    \caption{Plots between $P_{Min\, Err}$ and $\theta_0+\theta_1$ for different values of $\alpha$ and $\beta$}
    \label{fig:images}
\end{figure}
\par\noindent
From Fig. 2, it can also be observed that the standard (unmodified) Schr\"odinger equation ($\alpha = 0$ and $\beta=0$) generates a plot with a continuous underlying topology and the second wavefunction comes into existence at $2\bar{L}=1$. The underlying topology remains continuous even if we include the quartic correction term, namely with $\beta \neq 0$. 
What significantly changes the underlying topology is the cubic correction term, namely $\alpha \ne 0$. This produces a discrete structure of space and gives a shelved architecture to the graph. Knowing that $\bar{L}=2 n \pi \alpha m W_0/\hbar$  with $n\in \mathbb{N} $, we first fix the scale parameter $\alpha$ to $\alpha = 0.15/(2\pi m W_0 /\hbar)$) and, subsequently, we check the effect at $\alpha_1=\alpha$, $\alpha_2=\alpha/2$, and $\alpha_3=\alpha/3$. It is clear from the graph that as $\alpha$ decreases, i.e., first from $\alpha$ to $\alpha/2$ and then to $\alpha/3$, these shelves converge as the unit of length is directly proportional to $\alpha$. Also, at $\alpha=0$, the shelves merge and form a continuous underlying structure, as expected.
\par\noindent
Observations similar to the unmodified case can be found when the quartic correction term, namely with $\beta \ne 0$, is considered, but with a subtle difference such that the graph is shifted towards the left, that is, the second wave function appears before $2\bar{L}=1$. Since the minimum length for the second wavefunction has an opposite relation with $s_{11}$ (see Appendix B), the graph moves towards $2\bar{L}=1$ ( unmodified case as the value of $\beta$ decreases). To simplify the calculation, and since $s_{11}$ and $\lambda$ are both of the order of $\beta$, we equate them. We also consider $c_{11}=1$. Therefore, as the value of $\beta$ decreases from $\beta_1=\beta$ to $ \beta_2=\beta/2$ and then to $\beta_3=\beta/3$, it can be clearly seen that initially the plot is dislocated to the left compared to the unmodified case, and as the value of $\beta$ decreases, the plot tends to converge toward the unmodified case, which is expected (here we have taken $\beta=0.3 \hbar^2/(m^2 W_0^2)$).\\
%
%
%
%
%
\par\noindent
In conclusion, in this work, we proposed a novel method to test the effects of quantum gravity on information theory.    This method is based on quantum state discrimination, which has been developed in the context of quantum information theory. This is the first application of this method to quantum gravity. Thus, we used quantum state discrimination to demonstrate that quantum gravity approaches with an underlying discrete topology fundamentally differ from quantum gravity approaches for which spacetime has a continuous underlying topology.  Consequently, the low-energy implications of approaches such as string theory (with continuous underlying topology)  on quantum state discrimination would be very different from the low-energy implications of approaches such as loop quantum gravity (with discrete underlying topology). However, any modification that does not change the topology will not affect the quantum state discrimination.  
These classes of models produce different types of corrections to the low-energy effective field theory describing simple quantum systems, and these corrections can be used to affect the quantum state discrimination in a different way. We note that even though quantum gravitational effects are expected to become important at the Planck scale, they can also become important at scales much larger than the Planck scale. Hence, it is generally possible to experimentally test the results of the present work using currently available technologies. 
As quantum state discrimination has been used to study noise-robust quantum networks \cite{a0},   noisy quantum neural networks \cite{a1},  quantitative wave-particle duality \cite{a2}, decomposable entanglement witness \cite{a3}, entanglement swapping \cite{a4} and no-signaling bound \cite{a5}, a modification to quantum state discrimination would also modify these studies. In other words, it would be interesting to precisely analyze how discrete topology can change their behavior.  
\par\noindent
Finally, it should be pointed out that the low-energy consequences of quantum gravity need not be limited to higher derivative effects, but can produce more fundamental corrections to quantum mechanics, such as higher-order interference  \cite{ref5}. These higher-order interference effects could even, in principle, be measured \cite{ref6}, and could help understand low-energy implications of quantum gravity beyond higher derivative corrections. Moreover, it has been argued that such effects could occur in non-linear optics \cite{ref7}. 
Actually, the use of non-linear optics as an analogue for quantum gravity is well known \cite{ref6a, ref6b}. Therefore, it would be interesting to investigate if such higher-order interference can also act as an analogue of quantum gravitational effects.

%
%
%
%
%
%
%
%
%
%
\section*{Acknowledgments}
%
%
\par\noindent
We would like to thank Jan Kolodynski for constructive discussions on the quantum state discrimination. 
%
%
%
%
 
%
%
%
%

%
%
%
%
\section{Appendix A}
%
%
%
\par\noindent
Here we shall analyze the effects of cubic correction  terms \cite{1d, 1d1, 1d2,  1d4} on the wavefunction and the underlying topology of spacetime. Thus, we start from the cubic correction term  of the form 
\begin{equation}
\frac{d^2\psi}{d x^2}+2i\alpha\hbar\frac{d^3\psi}{d x^3}-k^2\psi(x)+\frac{2mW_0[\delta(x+L)+\delta(x-L)]}{\hbar^2}\psi(x)=0
\label{t6}
\end{equation}
with $k\equiv \sqrt{{-2mE}/{\hbar^2}} $.
Solving the above equation for $V(x)=0$, we get
$ 
\psi(x)=Ae^{i\gamma x}+Be^{k_1 x}+Ce^{-k_2x}
$
where $A$, $B$, $C$ are constants, and
$ 
\gamma\equiv {1}/{2\alpha\hbar},\quad k_1 \equiv k(1-i\alpha\hbar k),\quad k_2 \equiv k(1+i\alpha\hbar k)
$. 
After imposing the bound state condition, i.e., $\psi(-\infty)=\psi(\infty)=0$, the wave-function looks as 
$\psi(x) = s_1e^{i\gamma x}+c_{11}e^{ k_1x}$ for $x< -L$, 
$\psi(x) = s_2e^{i\gamma x}+c_{21}e^{ k_1x} + c_{22}e^{-k_2x}$ for $x\in (-L,L)$ and 
$\psi(x) = s_3e^{i\gamma x}+c_{32}e^{-k_2x}$ for $x> L$
where $s_1$, $s_2$, $s_3$, $c_{11}$, $c_{21}$, $c_{22}$, and $c_{32}$ are all constants of integration. We restrict  $c_{11}$ to take real values (by eliminating any imaginary phase associated with it),  and simplify further calculations. 
We can first use suitable boundary conditions to fix the arbitrary constants in the expression for the wavefunctions. Thus, using the continuity of wavefunctions at $x=-L$ we get
\begin{equation}
(s_1-s_2)e^{-i\gamma L} + (c_{11}-c_{21})e^{-k_1 L}- c_{22}e^{k_2L}=0~.
\label{t11}
\end{equation}
Similarly, at $x=+L$, we get
\begin{equation}
(s_2-s_3)e^{i\gamma L} + c_{21}e^{k_1 L}+ (c_{22}-c_{32})e^{-k_2L}=0~.
\label{t12}
\end{equation}
Now, we analyze the first jump condition. So, 
integrating Eq. (\ref{t6}) from $x=\pm L^-$ to $x=\pm L^+$, we get
\begin{equation}
\psi'(x)\bigg|^{\pm L^+}_{\pm L^-}+2i\alpha\hbar\psi''(x)\bigg|^{\pm L^+}_{\pm L^-}+\frac{2mW_0}{\hbar^2}\psi(\pm L)=0~.
\label{t13}
\end{equation}
At $x=-L$, after finding $\psi'(x)$ and $\psi''(x)$, and using the smallness of $\alpha$ in Eq. (\ref{t13}), we obtain
\begin{equation}
\frac{2mW_0}{\hbar^2}s_1e^{-i\gamma L}+\bigg( c_{21}k + \bigg(\frac{2mW_0}{\hbar^2}-k\bigg)c_{11} \bigg)e^{-k_1L}-c_{22}ke^{k_2L}=0~.
\label{t14}
\end{equation}
Similarly, at $x=+L$, from Eq. (\ref{t13}), we get
\begin{equation}
\frac{2mW_0}{\hbar^2}s_3e^{i\gamma L}-c_{21}ke^{k_1L}+\bigg( c_{22}k + \bigg(\frac{2mW_0}{\hbar^2} -k \bigg)c_{32} \bigg)e^{-k_2L}=0~.
\label{t16}
\end{equation}
\par\noindent
Thus, multiplying Eq. (\ref{t11}) by $k$ and subtracting it from Eq. (\ref{t14}), we obtain
$$\bigg[\bigg(1- \frac{2mW_0}{\hbar^2k}\bigg) s_1 - s_2 \bigg] e^{(k_1-i\gamma) L}+2\bigg[\bigg(1- \frac{mW_0}{\hbar^2k}\bigg) c_{11} - c_{21} \bigg]=0$$
which leads to
\begin{equation}
s_2=\bigg( 1-\frac{2mW_0}{\hbar^2k} \bigg) s_1 , \quad
c_{21}=\bigg( 1-\frac{mW_0}{\hbar^2k} \bigg) c_{11}~.
\label{t19}
\end{equation}
\par\noindent
These equations are a result of the assumption that $c_{11}$ and $c_{21}$ are deformed due to $\alpha$ in the same way, i.e., $\frac{c_{11}(\alpha)}{c_{21}(\alpha)}=\frac{c_{11}(\alpha=0)}{c_{21}(\alpha=0)}$. This is quite intuitive as both $c_{11}$ and $c_{21}$ are coefficients of $e^{k_1x}$.
Thus, multiplying Eq. (\ref{t12}) by $k$ and adding it to Eq. (\ref{t16}), we get
$$
\bigg[\bigg(1- \frac{2mW_0}{\hbar^2k}\bigg) s_3 - s_2 \bigg] e^{(k_2+i\gamma) L}+2\bigg[\bigg(1- \frac{mW_0}{\hbar^2k}\bigg) c_{32} - c_{22} \bigg]=0
$$
which leads to
\begin{equation}
s_2=\bigg( 1-\frac{2mW_0}{\hbar^2k} \bigg) s_3 , 
\quad
c_{22}=\bigg( 1-\frac{mW_0}{\hbar^2k} \bigg) c_{32}~.
\label{t22}
\end{equation}
Akin to how Eq. (\ref{t19}) is obtained, here a similar assumption is made regarding $c_{32}$ and $c_{22}$, i.e., $\frac{c_{32}(\alpha)}{c_{22}(\alpha)}=\frac{c_{32}(\alpha=0)}{c_{22}(\alpha=0)}$, as both $c_{32}$ and $c_{22}$ are coefficients of $e^{-k_2x}$.  \\
Hence, from Eqs. (\ref{t19}) and (\ref{t22}), we get
$
s_3=s_1
$. 
Now, using Eqs. (\ref{t19}) and (\ref{t22}) in  Eq. (\ref{t11}), we get
\begin{equation}
\frac{2mW_0}{\hbar^2k} s_1 e^{-i\gamma L}+ \frac{mW_0}{\hbar^2k}c_{11}e^{-k_1L}=\bigg( 1-\frac{mW_0}{\hbar^2k} \bigg) c_{32} e^{k_2L}~.
\notag
\end{equation}
Also, knowing $s_3=s_1$ and using Eqs. (\ref{t19}) and (\ref{t22}) in Eq. (\ref{t12}), we get
\begin{equation}
\frac{2mW_0}{\hbar^2k}s_1 e^{i\gamma L} + \frac{mW_0}{\hbar^2k}c_{32}e^{-k_2L}=\bigg( 1-\frac{mW_0}{\hbar^2k} \bigg) c_{11} e^{k_1L}~.
\notag
\end{equation}
Dividing the above two equations with each other and using the smallness of $\alpha$, we have
\begin{equation}
\frac{2 s_1 e^{(k_1-i\gamma) L}+ c_{11}}{2s_1 e^{(k_2+i\gamma) L} + c_{32}}=\frac{c_{32}}{c_{11}}~.
\notag
\end{equation}
Let $A_1\equiv e^{(k_1-i\gamma) L}$ and $A_2\equiv  e^{(k_2+i\gamma) L}$, and using them in the above equation, we get ${c_{32}}^2+(2A_2s_1)c_{32}-(2A_1 s_1 c_{11}+c_{11}^2)=0$. Thus, this equation is quadratic in $c_{12}$, so after solving it, we get
$
c_{12}=(-2A_2s_1\pm\sqrt{(2A_2s_1)^2+4(2A_1s_1c_{11}+c_{11}^2)})/2~.
$\\
Now, since $s_i$ is small, it is legitimate to consider $s_1^2 \approx 0$ and, hence, we have $c_{32}=-A_2s_1\pm c_{11}(1+2A_1 s_1 / c_{11} )^{1/2}$.
Consequently, after using Taylor expansion of the second term and with $s_1^2\approx0$, we get
$
c_{32}=\pm (c_{11}+(A_1 \mp A_2)s_1)
$.\\
Therefore, utilizing it in Eq. (\ref{t22}) to eliminate $c_{32}$, we get
$c_{22}=\pm (1-mW_0/\hbar^2k)(c_{11}+(A_1 \mp A_2)s_1)$.
Using now this result and Eq. (\ref{t19}) in Eq. (\ref{t14}) to eliminate $c_{22}$ and $c_{21}$, we obtain
\begin{equation}
\frac{2mW_0}{\hbar^2}s_1e^{-i\gamma L}+\bigg( \bigg( 1-\frac{mW_0}{\hbar^2k} \bigg) c_{11}k + \bigg(\frac{2mW_0}{\hbar^2}-k\bigg)c_{11} \bigg)e^{-k_1L}-(\pm (1-\frac{mW_0}{\hbar^2k})(c_{11}+(A_1 \mp A_2)s_1))ke^{k_2L}=0\nonumber~.
\end{equation}
Rearranging the terms of the above equation, the equation now reads
\begin{equation} 
\bigg( \frac{2mW_0}{\hbar^2} e^{-i\gamma L} +(A_2\mp A_1)(1-\frac{mW_0}{\hbar^2 k})k e^{k_2L}) \biggr) \frac{s_1}{c_{11}k} + \biggr( \frac{mW_0}{\hbar^2 k} e^{-k_1L} \mp (1-\frac{mW_0}{\hbar^2k}) e^{k_2L}\bigg)=0\nonumber~.
\end{equation}
Thus, we get
\begin{equation}
\pm\frac{s_1}{c_{11}} \frac{\eta}{k}e^{-k_2L} + \frac{mW_0}{\hbar^2 k}(1\pm e^{-2kL})=1 ,
\quad \text{where} \quad
\eta\equiv \frac{2mW_0}{\hbar^2}e^{-i\gamma L}+ (A_2\mp A_1)\bigg( 1-\frac{mW_0}{\hbar^2k} \bigg)k e^{k_2L}~.
\label{t30}
\end{equation}
At this point, we introduce the quantities
\begin{eqnarray}
&&\bar{L}\equiv \frac{mW_0L}{\hbar^2}, 
\quad \quad 
\bar{\gamma}\equiv \gamma L=\frac{L}{2\alpha\hbar},
\quad\quad 
\bar{k}\equiv kL, 
\quad\quad 
\bar{k_1}\equiv k_1L,
\quad\quad 
\bar{k_2}\equiv k_2L~.
\label{t32}
\end{eqnarray}
Now, as $s_1\alpha\approx0$, we can easily see that
\begin{equation}
s_1e^{\pm c\alpha}=s_1(1\pm c\alpha ...)\approx s_1~.
\label{t27}
\end{equation}
Thus, using Eq. (\ref{t27}), we have
\begin{equation}
\frac{1}{\bar{\gamma}^2}\approx 0\approx\frac{s_1}{\bar{\gamma}}, \quad s_1 e^{\bar{k_1}} = s_1 e^{\bar{k}}e^{-i\alpha\hbar k^2L}\approx s_1 e^{\bar{k}},\quad s_1 e^{\bar{k_2}} \approx  s_1 e^{\bar{k}}~.
\label{t33}
\end{equation}
Employing Eq. (\ref{t32}) in Eq. (\ref{t30}), we get
\begin{equation}
\pm\frac{s_1}{c_{11}} \frac{\eta}{k}e^{-\bar{k_2}} + \frac{\bar{L}}{\bar{k}}(1\pm e^{-2\bar{k}})=1~.
\label{t34}
\end{equation}
Multiplying $\eta$ with $\frac{s_1}{c_{11}}\frac{e^{-\bar{k_2}}}{k}$ and using  Eq. (\ref{t27}), we get
\begin{equation}
\frac{s_1}{c_{11}} \frac{\eta}{k}e^{-k_2L} = \frac{s_1}{c_{11}} \frac{2mW_0}{\hbar^2k} e^{-(k+i\gamma)L}  \mp\frac{s_1}{c_{11}}(A_1\mp A_2) \bigg( 1-\frac{mW_0}{\hbar^2k} \bigg)~.
\label{t35}
\end{equation}
Utilizing the values of $A_1$ and $A_2$, Eqs. (\ref{t32}) and (\ref{t33}), and the smallness of $\alpha$, we have 
$
s_1(A_1\mp A_2)=s_1 (e^{-i\bar{\gamma}}\mp e^{i\bar{\gamma}})e^{\bar{k}}
$. 
Thus, combining this equation with Eqs. (\ref{t32}) and (\ref{t33}) in Eq. (\ref{t35}), we get
$$
\frac{s_1}{c_{11}} \frac{\eta}{k}e^{-k_2L} =\frac{s_1}{c_{11}}\bigg[ \frac{2\bar{L}}{\bar{k}}e^{-(\bar{k}+i\bar{\gamma})} + \bigg(1-\frac{\bar{L}}{\bar{k}}\bigg)e^{\bar{k}} (e^{i\bar{\gamma}}\mp e^{-i\bar{\gamma}}) \bigg]~.
$$
Now, using the above equation in Eq.  (\ref{t34}), we get
\begin{equation}
\frac{\bar{L}}{\bar{k}}=\frac{1 \mp \frac{s_1}{c_{11}}\bigg[ \frac{2\bar{L}}{\bar{k}}e^{-(\bar{k}+i\bar{\gamma})} + \bigg(1-\frac{\bar{L}}{\bar{k}}\bigg)e^{\bar{k}} (e^{i\bar{\gamma}}\mp e^{-i\bar{\gamma}}) \bigg]}{(1\pm e^{-2\bar{k}})}~.
\label{t38}
\end{equation}
For the wavefunction with the upper sign, we now get
\begin{equation}
\frac{\bar{L}_u}{\bar{k}}=\frac{1 - \frac{s_1}{c_{11}}\bigg[ \frac{2\bar{L}_u}{\bar{k}}e^{-(\bar{k}+i\bar{\gamma})} + \bigg(1-\frac{\bar{L}_u}{\bar{k}}\bigg)e^{\bar{k}} (e^{i\bar{\gamma}}- e^{i\bar{\gamma}}) \bigg]}{(1+ e^{-2\bar{k}})}\nonumber~.
\end{equation}
Using $e^{-i\gamma} = \cos(\gamma) - i \sin(\gamma)$ and $e^{i\gamma} - e^{-i\gamma} = 2i \sin(\gamma)$, we get
\begin{equation}
\frac{\bar{L}_u}{\bar{k}} = \frac{1- \frac{S_1}{C_{11}}(\frac{2\bar{L}_ue^{-\bar{k}}}{\bar{k}}(\cos(\gamma)-i \sin(\gamma)) + (1 - \frac{\bar{L}_u}{\bar{k}})e^{\bar{k}}(2 i\sin(\gamma)}{1+e^{2\bar{k}} }~. \nonumber
\end{equation}
Finally, we obtain
\begin{equation}
\frac{\bar{L}_u}{\bar{k}}=\frac{1 - \frac{2s_1}{c_{11}}\bigg[ \frac{\bar{L}_u}{\bar{k}}e^{-\bar{k}} \cos{\bar{\gamma}} + i \bigg(1-\frac{\bar{L}_u}{\bar{k}}(1+e^{-2\bar{k}})\bigg)e^{\bar{k}}  \sin{\bar{\gamma}}\bigg]}{(1 + e^{-2\bar{k}})}~.
\label{t39}
\end{equation}
Similarly, for the wavefunction with the lower sign in Eq. (\ref{t38}), we obtain
\begin{equation}
\frac{\bar{L}_l}{\bar{k}}=\frac{1 + \frac{2s_1}{c_{11}}\bigg[  \bigg(1-\frac{\bar{L}_l}{\bar{k}}(1-e^{-2\bar{k}})\bigg)e^{\bar{k}} \cos{\bar{\gamma}} - i\frac{\bar{L}_l}{\bar{k}} e^{-\bar{k}}  \sin{\bar{\gamma}}\bigg]}{(1 - e^{-2\bar{k}})}~.
\label{t40}
\end{equation}
%
%
%
%
%
%
%
%
%
%
\par\noindent
We keep $c_{11}$ real by taking out any imaginary phase and we assume $s_1\in \mathbb{R}$. 
Hence, equating the imaginary parts of Eq. (\ref{t39}), we obtain
\begin{equation}
\bigg(1-\frac{\bar{L}_u}{\bar{k}}(1+e^{-2\bar{k}})\bigg)e^{\bar{k}}  \sin{\bar{\gamma}}=0\nonumber
\end{equation}
which tells us either 
$
\frac{\bar{L}_u}{\bar{k}}=\frac{1}{(1 + e^{-2\bar{k}})},
$ 
or 
$
\sin{\bar{\gamma}}=0
$. 
As the former one is in contradiction to Eq. (\ref{t39}), since $s_1\neq 0$ in general, we therefore accept the latter one.
Similarly, equating the imaginary parts of Eq. (\ref{t40}),  we also obtain
$
\sin{\bar{\gamma}}=0
$.
Therefore, for both Eqs. (\ref{t39}) and (\ref{t40}), and with $s_1\in \mathbb{R}$, we have
\begin{equation}
\sin{\bar{\gamma}}=0 \implies \bar{\gamma}=n\pi \quad \forall\, n\in \mathbb{N} \implies L=2n\pi\alpha\hbar \quad \text{or} \quad \bar{L}=n\alpha \frac{2\pi m W_0}{\hbar}~.
\label{t41}
\end{equation}
Therefore, Eq. (\ref{t41}) shows the discrete nature of space by demonstrating the discreteness of the length $L$ between the two delta-wells.
%
%
%
%
%
%
%
%
%
%
\par\noindent
Now, let us see how the two wavefunctions come into being near 
$\bar{k}\to 0$.
For $\bar{k}\to 0$, the condition for the wavefunction described by Eq. (\ref{t39}) looks like 
$
\bar{L}_{u0}= -(s_1/c_{11})\bar{L}_{u0}(\cos{\bar{\gamma}}-2i\sin{\bar{\gamma}})
$, 
where, $\bar{L}_{u0}$ depicts the minimum value of $\bar{L}_u$ for the wavefunction described by Eq. (\ref{t39}).
Thus, from above equation, by comparing the real parts, we see that $s_1$ and $c_{11}$ have opposite signs as we have assumed that $s_1$ is real.
Solving this equation gives 
$
\bar{L}_{u0} ( 1+(s_1/c_{11})(\cos{\bar{\gamma}}-2i\sin{\bar{\gamma}}) ) =0
$.
Thus, either $\bar{L}_{u0}=0$, or $c_{11}= - s_1 (\cos{\bar{\gamma}}-2i\sin{\bar{\gamma}})$. 
The latter one cannot be the case, as this implies that if $\alpha=0$, $c_{11}=0$ (as here $s_1=0$). Therefore, we remain with the only choice that 
$
\bar{L}_{u0}=0.
$
Therefore, this wavefunction gets to exist from the beginning, i.e., from $L=0$.
\par\noindent
Now, for $\bar{k}\to 0$, the condition for the wavefunction described by Eq. (\ref{t40}) takes the form
\begin{equation}
\bar{L}_{l0}= \frac{1}{2} + \lim_{\bar{k}\to0}\frac{s_1}{c_{11}} \bigg[(1-2\bar{L}_{l0})\cos{\bar{\gamma}}- i\frac{\bar{L}_{l0} }{\bar{k}} \sin{\bar{\gamma}} \bigg]
\nonumber
\end{equation} 
where, $\bar{L}_{l0}$ depicts the minimum value of $\bar{L}_l$ for which the wavefunction described by Eq. (\ref{t40}) exists.
This equation can only be plausible if $\sin{\bar{\gamma}}=0 \implies\bar{\gamma}=n\pi$, which is what we obtained in Eq. (\ref{t41}). Hence, the assumption of considering $s_1$ as real perfectly works.
Thus, 
$
(s_1/c_{11}) (2\bar{L}_{l0}-1)\cos{\bar{\gamma}} \implies (2\bar{L}_{l0}-1)(1+2(s_1/c_{11})\cos{\bar{\gamma}})=0. 
$
Hence, as $s_1/c_{11}=-1/(2\cos{\bar{\gamma}})$ is not feasible, as $\cos{\bar{\gamma}} = (-1)^n$ and $s_1$ is very near to zero unlike $c_{11}$, we thus remain with only one option, i.e., 
$
2\bar{L}_{l0}-1=0 \implies 2\bar{L}_{l0}=1
$. \\
Therefore, if we consider $L$ to be continuous, then it can be said that the condition for the second bound state to exist is given by 
$
2\bar{L}\ge 1 \implies L\ge \hbar^2/(2mW_0)
$. 
This is quite intuitive, as the closer the wells will get, they will seem to act like a single well, i.e., have a single wavefunction.
Thus, considering the above equation and employing Eq.  (\ref{t41}) in it gives the following condition for observing the second bound state in the discrete space
\begin{equation}
n \ge \bigg\lceil{\frac{\hbar}{4\pi\alpha m W_0}} \bigg\rceil \implies L \ge 2\pi\alpha\hbar\bigg\lceil{\frac{\hbar}{4\pi\alpha m W_0}} \bigg\rceil~.
\label{t45}
\end{equation}
We took the least integer function here as $n$ is discrete.
Thus, the above equation gives the minimum value of $L$, namely the length between the two delta wells, for the existence of two bound states. \\
Now, using Eq. (\ref{t41}) in Eqs. (\ref{t39}) and (\ref{t40}) to replace cosine and sine, we get
\begin{equation}
\frac{\bar{L}_u}{\bar{k}}=\frac{1 - \frac{2s_1}{c_{11}} \frac{\bar{L}_u}{\bar{k}}e^{-\bar{k}}(-1)^n  }{(1 + e^{-2\bar{k}})}
\quad \text{and} \quad
\frac{\bar{L}_l}{\bar{k}}=\frac{1 + \frac{2s_1}{c_{11}}  \bigg(1-\frac{\bar{L}_l}{\bar{k}}(1-e^{-2k})\bigg)e^{\bar{k}} (-1)^n}{(1 - e^{-2\bar{k}})}
\label{t47}~.
\end{equation}
The second terms in the numerators of both the above equations represent the correction due to the presence of the scale parameter $\alpha$.
%
%
%
%
%
%
%
\section{Appendix B}
%
%
%
\par\noindent
Here we shall analyze the effects of  quartic  correction terms \cite{1c, 1c1, 1c2,  1c4} on the wavefunction and the underlying topology of spacetime. Thus, we start with the quartic correction term of the form 
\begin{equation}
\biggr(  -\frac{\hbar^2}{2m}\frac{d^2}{dx^2}+\frac{\beta \hbar^4}{m}\frac{d^4}{dx^4}-[W_o \delta (x+L) + W_o \delta (x-L) + E] \biggl)\psi(x)=0
\label{g3}~.
\end{equation}
We can express $\eta$ as 
\begin{equation}
\eta\equiv\frac{1}{\hbar\sqrt{2\beta}}~.
\label{g4}
\end{equation}
Now, as $V(x)=0$ for $x\ne \pm L$,  we obtain 
$
\frac{d^2 \psi(x)}{dx^2}-2\beta\hbar^2\frac{d^4 \psi(x)}{dx^4}- k^2 \psi(x)=0
$~.
Hence, solving this equation,  we  obtain  
$
\psi(x)=Ce^{\eta x}+D^{-\eta x}+Ae^{bx}+Be^{-bx}
$ with $b\equiv k+\hbar^2k^3\beta$~.
\par\noindent
Hence, from the above equations, after utilizing the bound state conditions  $\psi(-\infty) = \psi(\infty ) = 0$, we get the complete equation for the double-delta potential well in the form
\begin{equation}
\psi(x) = \begin{cases}
					s_{11}e^{\eta x}+c_{11}e^{ bx}, & x< -L\\
					s_{21}e^{\eta x}+s_{22}e^{-\eta x}+c_{21}e^{ bx} + c_{22}e^{-bx}, & x\in (-L,L)\\
					s_{32}e^{-\eta x}+c_{32}e^{-bx}, &  x>L
		  \end{cases}~.
\label{g42}
\end{equation}
It is important to note that the impact of terms with $\eta$ in them is small, so $s_{11},\, s_{21},\, s_{22},\, s_{32}$ are all very small because if $\beta \to 0$, then these terms must get killed. Thus,  we can write 
$
s_{ij}\beta^m= 0=s_{ij}^n\quad \forall\,i,j$ with
$m\ge 1$ and $ n>1
$. 
Now, since the potential is symmetric here, so even-odd parity can be used, i.e., $\psi(x)=\pm\psi(-x)$. 
Thus, using this in the region $||x|| > 0$, we have $\psi(-x:x<-L)=\pm\psi(x: x>L)$ which gives, $(s_{11} \mp s_{32}) = -(c_{11} \mp c_{32}) e^{(\eta-b) x} $. Hence, knowing here that $x$ is a variable, it holds only when
\begin{equation}
s_{11} = \pm s_{32},\, c_{11}=\pm c_{32}
\label{g74}~.
\end{equation}
Now, using  the even-odd parity in the region $||x|| < 0$, we get $\psi(-x:-L<x<0)=\pm\psi(x: 0<x<L)$ which  leads to
\begin{equation}
     (s_{22} \mp s_{21}) = -(c_{22} \mp c_{21}) \frac{(e^{bx} \mp e^{-bx})}{(e^{\eta x} \mp e^{-\eta x})}
\label{g73}~.
\end{equation}
Now, differentiating this expression, we obtain 
\begin{equation}
\psi^{(n)}(x) = \begin{cases}
					s_{11}\eta^n e^{\eta x}+c_{11}b^n e^{ bx}, & x< -L\\
					s_{21}\eta^n e^{\eta x}+(-1)^n s_{22}\eta^n e^{-\eta x}+c_{21}b^n e^{ bx} + (-1)^n c_{22}b^n e^{-bx}, & x\in (-L,L)\\
					s_{32}\eta^n e^{-\eta x}+(-1)^n c_{32}b^n e^{-bx}, &  x>L
		  \end{cases}~.
\label{g48}
\end{equation}
Also, by rearranging Eq. (\ref{g3}), we get
\begin{equation}
\psi''(x)-2\beta\hbar^2\psi''''(x)+\bigg[ \frac{2mW_0}{\hbar^2}[\delta(x+L)+\delta(x-L)]-k^2 \bigg]\psi(x)=0
\label{g49}~.
\end{equation}
After integrating  from $x=\pm L^-$ to $x=\pm L^+$, we obtain the jump condition
\begin{equation}
\psi'(x) \bigg|^{\pm L^+}_{\pm L^-}-2\beta\hbar^2\psi'''(x)\bigg|^{\pm L^+}_{\pm L^-}= - \frac{2mW_0}{\hbar^2}\psi(\pm L)
\label{g50}~.
\end{equation}
Taking the lower sign in the above equation, we get
\begin{equation}
[Q(c_{21}-c_{11})-c_{11}]e^{-bL}-Qc_{22}e^{bl}-s_{11}e^{-\eta L}=0
\label{g51}
\end{equation}
while solving with the upper sign, we obtain
\begin{equation}
[Q(c_{22}-c_{32})-c_{32}]e^{-bL}-Qc_{21}e^{bl}-s_{32}e^{-\eta L}=0
\label{g53}
\end{equation}
where
\begin{equation}
Q=-\frac{\hbar^2b(1-2\beta\hbar^2b^2)}{2mW_0}
\label{g52}~.
\end{equation}
Thus,  we get 
$
    (c_{22}\mp c_{21})(e^{-bl}\mp e^{bl})=0 \implies c_{22} =\pm c_{21}
$~.
Using this result in Eq. (\ref{g73}), we can also obtain   
$
    s_{22} =\pm s_{21}~.
$
Therefore, summarizing the results,  we have
\begin{equation}
s_{21}=\pm s_{22},\,s_{32}=\pm s_{11},\,c_{21}=\pm c_{22},\,c_{32}=\pm c_{11}~.
\label{g58}
\end{equation}
Now, double integrating   from $x=\pm L^-$ to $x=\pm L^+$ with respect to $x$, we get the second jump condition
\begin{equation}
\psi(x)\biggr|^{\pm L^+}_{x=\pm L^-}=2\beta\hbar^2 \psi''(x)\biggr|^{\pm L^+}_{x=\pm L^-}~.
\end{equation}
Thus,  we obtain the following result for both $+L$ and $-L$
\begin{equation}
c_{22}=\frac{c_{11}}{e^{2bL}\pm 1}
\label{g60}~.
\end{equation}
Also,  we  can write
\begin{equation}
\biggr[Q(\frac{c_{11}}{e^{2bL}\pm 1}\mp c_{11})\mp c_{11}\biggr]e^{-bL}\mp Q\frac{c_{11}}{e^{2bL}\pm 1}e^{bl}\mp s_{11}e^{-\eta L}=0~.
\notag
\end{equation}
At this point, we introduce the parameters  
\begin{equation}
\bar{L}\equiv \frac{mW_0L}{\hbar^2},\quad \lambda \equiv \frac{m^2W_0^2\beta}{\hbar^2}=\frac{1}{2}\bigg(\frac{ \bar{L}}{\eta L}\bigg)^2,  \quad \bar{k}=kL, \quad \bar{b} \equiv bL=\bar{k} + \lambda \frac{\bar{k}^3}{\bar{L}^2}~.
\label{g27}
\end{equation}
Thus, we obtain 
$
-2Q=(\frac{s_{11}}{c_{11}}e^{-\frac{ \bar{L}}{\sqrt{2\lambda}}+ \bar{b}} +1)(1\pm e^{-2 \bar{b}})
$, and 
$
Q=-\frac{1}{2}\frac{ \bar{b}}{ \bar{L}}( 1-2\lambda\frac{ \bar{b}^2}{ \bar{L}^2} )
$. 
Therefore, from these two expressions, we obtain
\begin{equation}
\frac{ \bar{L}}{ \bar{b}}=\frac{\biggr( 1-2\lambda\frac{ \bar{b}^2}{ \bar{L}^2} \biggr)}{\bigg(\frac{s_{11}}{c_{11}}e^{-\frac{ \bar{L}}{\sqrt{2\lambda}}+ \bar{b}}+1\bigg)}  \frac{1}{(1\pm e^{-2 \bar{b}})}~.
\label{g63}
\end{equation}
The above equation gives the impression that this system has two wavefunctions, as expected. Also, from this equation, we obtain  $s_{11}/c_{11} \in \mathbb{R} $, as the exponential multiplied with it cannot make the term to vanish if it is imaginary. Furthermore, it is assumed $\frac{s_{11}}{c_{11}}>0$, since if this quantity is negative and the exponential takes the value that makes the term equal to $-1$, then this term will blow up the expression, which should not be the case.
\par\noindent
Now, let us find the minimum length between the wells necessary for the existence of a particular wavefunction. Since $\bar{b} =\bar{k} + \lambda \frac{\bar{k}^3}{\bar{L}^2}$, it is obvious that  if $\bar{k}\to 0$, then $ \bar{b}\to 0$. Therefore,
\begin{equation}
 \bar{L}_{ \bar{b}=0}=
\begin{cases}
0 \quad\quad\quad\quad\quad\quad\quad\quad\quad\, \text{for even wave-function} \\  
\frac{1}{2}\bigg(1-\frac{s_{11}}{c_{11}}e^{-\frac{ \bar{L}_{ \bar{b}=0}}{\sqrt{2\lambda}}}\bigg) \quad \text{for odd wave-function}
\end{cases}~.
\label{g64}
\end{equation} 
\begin{figure}[h]
\includegraphics[scale=0.45]
{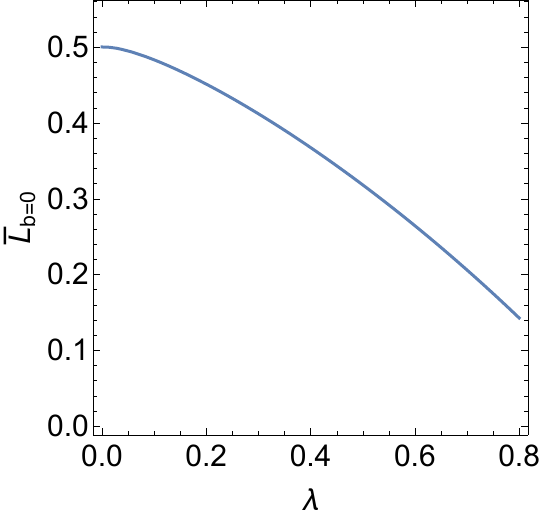}
\\

\caption{The plot here is $\lambda $ vs $\bar{L}_{b=0}$ which is obtained by plotting Eq. (\ref{g64}) for odd wavefunction while assuming $C_{11} \sim 1$ and $S_{11} \sim \lambda$ since both are of the order of $\beta$. }
\end{figure}
\par\noindent
In Fig. 3, the plot clearly shows that as $\lambda$ or, equivalently, $\beta$ decreases, $\bar{L}$ increases and, finally, it saturates to the value $1/2$ when $\lambda=0$ or, equivalently, $\beta=0$) which is what is anticipated.
\par\noindent
Now, simplifying Eq. (\ref{g63}) and taking into consideration the smallness of $s_{11}$ and $\lambda$,  we obtain
\begin{equation}
    \bar{L}=\bigg(1-\frac{s_{11}}{c_{11}}e^{-\frac{ \bar{L}}{\sqrt{2\lambda}}+ \bar{k} + \lambda \frac{\bar{k}^3}{\bar{L}^2}}\bigg)   \frac{ \bar{k}-\lambda\frac{ \bar{k}^3}{ \bar{L}^2} }{(1\pm e^{-2 (\bar{k} + \lambda \frac{\bar{k}^3}{\bar{L}^2})})} 
    =   \frac{ \bar{k}-\lambda\frac{ \bar{k}^3}{ \bar{L}^2} - \frac{s_{11}}{c_{11}} \bar{k} e^{-\frac{ \bar{L}}{\sqrt{2\lambda}}+ \bar{k} + \lambda \frac{\bar{k}^3}{\bar{L}^2}} }{(1\pm e^{-2 (\bar{k} + \lambda \frac{\bar{k}^3}{\bar{L}^2})})}
    \label{g71}~.
\end{equation}
%
%
%
%
%
%
Therefore, unlike the cubic correction terms, the underlying topology is continuous here. However, the quartic correction term explicitly induces a modification in the wavefunction.

%
%
%
%
%
%
%
%
%
%
%
%
%
%
%
%
%
%
%
%
%
%
%
%
%
%
%
%

\end{document}